\documentclass[12pt]{article}
\usepackage{graphicx}
\usepackage{subfigure}

\begin{document}

\begin{center}
\begin{large}
\title\\{\textbf{An analysis of the Isgur-Wise Function and its derivatives within a Heavy-Light QCD Quark Model}}\\\

\end{large}

\author\

\textbf{$Sabyasachi\;Roy^{\emph{1*}}\;and\:D\:K\:Choudhury^{\emph{1,2,3}}$} \\\

1. Dept. of Physics, Gauhati University, Guwahati-781014, India.\\
2. Centre for Theoretical Studies, Dept. of Physics, Pandu College,Guwahati-781012, India.\\
3. Physics Academy of the North East, Guwahati-781014, India. \\

* On leave from Karimganj College, Karimganj, India-788710 ; e-mail :  \emph{sroy.phys@gmail.com}

\begin{abstract}

In determining the mesonic wave function from QCD inspired potential model, if the linear confinement term is taken as parent ( with columbic term as perturbation ), Airy's function appears in the resultant wave function - which is an infinite series. In the study of Isgur-Wise function (IWF) and its derivatives with such a wave function, the infinite upper limit of integration gives rise to divergence. In this paper, we have proposed some reasonable cut-off values for the upper limit of such integrations and studied the subsequent effect on the results. We also study the sensitivity of the order of polynomial approximation of the infinite Airy series in calculating the derivatives of IWF.
\begin{flushleft}
\end{flushleft}
\end{abstract}
\end{center}

\section{Introduction:}\rm
\ \ \ The quark composite systems with one heavy quark have been the focus of interest for the last few decades. In infinite heavy quark mass limit, the heavy quark sector of QCD becomes independent of quark masses and the effective Lagrangian of the heavy quark effective theory ( HQET ) exhibits additional spin flavour symmetries and this simplifies the calculations of matrix elements of electroweak transitions.
In semi-leptonic transitions,in the limit of infinite quark masses, all the mesonic form factors can be expressed in terms of a single universal function, called the Isgur-Wise Function [1]. The shape of the IWF and its derivatives (slope and curvature) at zero recoil, is essential for determination of CKM matrix elements and this requires a reasonable description of the IWF.
The main part of the IWF is the wave function of the hadrons. The wave function for the heavy-light mesons have been calculated earlier within the framework of QCD potential model  [2,3] with considerable accuracy [4]. This has been deduced both with columbic term in potential as parent [5] and also with the linear confinement term as parent [6]. The characteristics like slope (charge radii) and curvature (convexity parameter) of IWF have been reported in both the two cases with certain limitations.
In the present paper, we have considered the linear confinement term  (br) as parent and Columbic term as perturbation. This has been done applying Dalgarno's method of perturbation theory up to first order correction [7]. As Airy's function appears in the wave function, the integrability of the otherwise divergent infinite Airy's function series in the IWF is a question of consideration in the present work.
We have introduced some reasonable cut-off for upper limit of integration of IWF keeping in mind the nature of Airy's function and the boundary condition of IWF ( $\xi$(1)$=1$) and studied the variation of the result with this cut-off value at different orders of Airy's function. We also study the sensitivity of the order of polynomial approximation of the infinite Airy's function when compared with experimental result of the derivatives of IWF.
The section 2 contains the essential formalism, calculations and results are reported in section 3 and the section 4 contains conclusion.

\section{Formalism:}
\subsection{Potential Model:}
The potential under consideration ( Cornell potential ) is :
\begin{equation}
V (r) = -\frac{4\alpha_s}{3r} + br + c
\end{equation}
Here we take  \textbf{br} as parent  so that our unperturbed Hamiltonian [8] is
\begin{equation}
H_0 = -\frac{\nabla^2}{2\mu}+br
\end{equation}
with
\begin{equation}
H^\prime = -\frac{4\alpha_s}{3r}+ c
\end{equation}
as perturbation. Here $\mu$ is the reduced mass, which is
\begin{equation}
   \mu= \frac{m_qm_Q}{m_q+m_Q}
\end{equation}
We take the value of b to be 0.183 $GeV^2$ from charomonium spectroscopy [9,10] and constant c to be 1 GeV [5].
It is to be mentioned that, in the infinite heavy quark mass limit ( $m_Q\rightarrow\infty$ ) ,
\begin{equation}
\mu = \lim_{m_Q \rightarrow \infty }\frac{m_q m_Q}{m_q + m_Q}\approx m_q
\end{equation}

Under this consideration, the two body Schrodinger equation [12, 13] for the Hamiltonian $ H = H_0 + H^\prime $ is :
\begin{equation}
H|\Psi>=(H_0+H^\prime)|\Psi>=E|\Psi>
\end{equation}
\subsection{Wave Function:}
To find the unperturbed wave function corresponding to $H_0$    we employ the radial Schrodinger equation for potential br for ground state S ($l=0$), following the formalism of reference [14], as :
\begin{equation}
[-\frac{1}{2\mu}(\frac{d^2}{dr^2}+\frac{2}{r}\frac{d}{dr})+br]R(r)=ER(r)
\end{equation}
where R(r) is the radial wave function.
We introduce u(r) = r R ( r ) and the dimensionless variable $\varrho(r)$ , where -
\begin{equation}
\varrho =(2\mu b)^{1/3}r-(\frac{2\mu}{b^2})^{1/3}E
\end{equation}
The equation (7) then reduces to:
\begin{equation}
\frac{d^2 u}{d\varrho^2}-\varrho u = 0
\end{equation}
The solution of this second order homogeneous differential equation [15] contains linear combination of two types of Airy's functions Ai[r] and Bi[r]. The nature of the Airy's function [16] reveals that \\
\begin{center}
$Ai[r]\rightarrow 0$ and  $ Bi[r]\rightarrow \infty$ as  $r\rightarrow \infty$.
\end{center}
So, it is reasonable to reject the Bi[r] part of the solution. The radial wave function has thus the form:
\begin{equation}
u(r)=NA_i[(2\mu b)^{1/3} (r-\frac{E}{b})]
\end{equation}
where N is our normalization constant    which has the dimension of $GeV^{1/2}$ .The boundary condition u(0) $=$ 0 [17] gives us the unperturbed energy for ground state [13]:
\begin{equation}
W^{0}=E=-(\frac{b^2}{2\mu})^{1/3} \varrho_0
\end{equation}
where $\varrho_0$ is the zero of the Airy function , such that Ai[$\varrho_0$]$=$0 [16].\\
$\varrho_0$ has the explicit form -
\begin{equation}
\varrho_0=-[\frac{3\pi(4n-1)}{8}]^{2/3}
\end{equation}
(In our case $n=1$ for ground state.) From this, we get the unperturbed wave function for ground state to be :

\begin{equation}
\Psi^0(r)=\frac{N}{2\sqrt{\pi}r}A_i[\varrho_1 r + \varrho_0]=\frac{N}{2\sqrt{\pi}r}A_i[\varrho]
\end{equation}
where we have taken $\varrho_1$ $=$ $(2\mu b)^{1/3}$ and $\varrho = \varrho_1 r + \varrho_0 $ .\\
The first order perturbed eigen function $\Psi^{\prime}$ and eigen energy $W^{\prime}$ can be calculated using the following relation:
\begin{center}
\begin{equation}
H_0\Psi^\prime + H^\prime\Psi^0 = W^{0} \Psi^\prime + W^{\prime}\Psi^0
\end{equation}
\end{center}

We find,
\begin{center}
\begin{equation}
W^{\prime}=\int_0^\infty r^2H^\prime \mid\ \Psi^0 \mid^ 2 dr
\end{equation}
\end{center}

Employing Dalgarno's method [18] , the first order wave function comes out to be [6]:
\begin{equation}
\Psi^{\prime}(r)= -\frac{4\alpha_s}{3}(\frac{a_0}{r}+a_1 r+a_2)
\end{equation}
where $a_0  ,a_1$  and $a_2$ are terms which involve $\alpha_s  ,b,\mu , W^{\prime}$, E and c. These are having dimensions of $GeV^{1/2}, GeV^{3/2} , GeV^{5/2}$ respectively and have explicit form, considering Airy order up to $r^3$, as given below [6].

\begin{center}
\begin{eqnarray}
a_0 = \frac{0.8808(b\mu)^{1/3}}{(E-c)}-\frac{a_2}{\mu (E-c)}+ \frac{4W^{\prime}\times0.21005}{3\alpha_s(E-c)} \\
a_1 = \frac{ba_0}{(E-c)} + \frac{4W^{\prime}\times0.8808(b\mu)^{1/3}}{3\alpha_s(E-c)}-\frac{0.6535\times(b\mu)^{2/3}}{(E-c)}\\
a_2 = \frac{4\mu W^{\prime}\times 0.1183}{3\alpha_s}
\end{eqnarray}
\end{center}

The total wave function with first order correction is:
\begin{equation}
\Psi_{tot} (r) = \Psi^0 (r) + \Psi^{\prime} (r)
\end{equation}
which upon substitution yields-
\begin{equation}
\Psi_{tot} (r) = \Psi^0 (r) + \Psi^{\prime}= N^\prime[\frac{1}{2\sqrt{\pi}r}A_i(\varrho_1 r + \varrho_0)- \frac{4\alpha_s}{3}(\frac{a_0}{r}+a_1 r+a_2)]
\end{equation}
Here $N^\prime $ is the normalization constant of total wave function which is also having the dimension of $GeV^{1/2}$.
Considering relativistic effect on the wave function, the total relativistic wave function is given by [12]:
\begin{equation}
\Psi_{rel} (r) = N^\prime[\frac{1}{2\sqrt{\pi}r}A_i(\varrho_1 r + \varrho_0)- \frac{4\alpha_s}{3}(\frac{a_0}{r}+a_1 r+a_2)](\frac{r}{a_b})^{-\epsilon }
\end{equation}
Here,
\begin{equation}
a_b = \frac{3}{4\mu \alpha_s} \;\; and\;\;  \epsilon = 1-\sqrt{1-(\frac{4\alpha_s}{3})^2}
\end{equation}
\subsection{Isgur-Wise Function:}
In case of semi-leptonic decay of hadrons ( mesons ) , in the infinite mass limit, a new symmetry called spin-flavored symmetry, will emerge and the Heavy Quark Effective Theory ( HQET ) will be suitable. In this theory, the strong interactions of the heavy quarks are independent of its spin and mass[19] and all the form factors are completely determined, at all momentum transfers, in terms of only one elastic form factor function, the universal Isgur-Wise function $\xi( v,v^\prime )$.  $\xi( v,v^\prime )$ depends only upon the four velocities   $v_\nu $ and $v_{\nu^\prime} $  of heavy particle before and after decay. This $\xi( v,v^\prime )$  is normalized at zero recoil [20].
If y = $v_\nu $.$v_{\nu^\prime} $ , then, for zero recoil  $\xi(1)=1$.
In explicit form IW function can be expressed as :
\begin{equation}
\xi(y)=1-\rho^2 (y-1) +C(y-1)^2 + ......
\end{equation}
$\rho^2$ is the slope parameter and is given by -
\begin{equation}
\rho^2 = -\frac{\delta\xi (y)}{\delta y}|_{y=1}
\end{equation}
\begin{flushright}
$\rho$ is known as the charge radius. \\
\end{flushright}
C is the convexity parameter given by -
\begin{equation}
C= \frac{\delta^{2}\xi (y)}{\delta y^2}|_{y=1}
\end{equation}
The calculation of this IWF is non-perturbative in principle and is performed for different phenomenological wave functions for mesons [21]. This function depends upon the meson wave function and some kinematic factor, as given below :
\begin{equation}
\xi(y)=\int_0 ^\infty 4\pi r^2 |\Psi(r)|^2\cos(pr)dr
\end{equation}
where $\cos(pr)=1-\frac{p^2 r^2}{2}+\frac{p^4 r^4}{24}$ +$\cdot\cdot\cdot\cdot\cdot\cdot$  with $ p^2=2\mu^2 (y-1).$ Taking cos(pr) up to  $O(r^4)$ we get,

\begin{flushleft}
\begin{eqnarray}
\xi(y)= \int_0 ^\infty 4\pi r^2 |\Psi(r)|^2dr - [4\pi\mu^2\int_0^\infty r^4|\Psi(r)|^2dr](y-1)+[\frac{2}{3}\pi\mu^4\int_0^\infty r^6|\Psi(r)|^2dr](y-1)^2
\end{eqnarray}
\end{flushleft}

Equations (25) and (29) give us :
\begin{eqnarray}
\rho^2 = [4\pi\mu^2\int_0^\infty r^4|\Psi(r)|^2dr] \\
C= [\frac{2}{3}\pi\mu^4\int_0^\infty r^6|\Psi(r)|^2dr] \;\;\;\; and \\
\int_0 ^\infty 4\pi r^2 |\Psi(r)|^2dr =1
\end{eqnarray}
Equation (32) gives the normalization constants $ N $ and $N^{\prime}$ for $\Psi^0 (r)$ and $\Psi_{tot} (r)$  respectively, as :
\begin{center}
\begin{eqnarray}
N=\frac{1}{(\int_0^\infty A_i[(\varrho_1 r + \varrho_0])^{1/2}} \\
N^{\prime} = \frac{1}{[4\pi\int_0^\infty r^2(\frac{1}{2\sqrt{\pi}r} A_i(\varrho_1 r + \varrho_0)- \frac{4\alpha_s}{3}(\frac{a_0}{r}+br+c)]^{1/2}}
\end{eqnarray}
\end{center}
\section{Calculation and result:}
With linear confinement term of potential as parent, the wave functions contain Airy's function $Ai[\varrho]$ , which is an infinite series in itself [23].
\begin{eqnarray}
A_i[\varrho] = a[1+\frac{\varrho^3}{6}+\frac{\varrho^6}{180}+\frac{\varrho^9}{12960}+...]-
 b[\varrho +\frac{\varrho^4}{12}+\frac{\varrho^7}{504}+\frac{\varrho^{10}}{45360}+...]
\end{eqnarray}

\begin{flushright}
 with $a=0.3550281, b=0.2588194.$ \\
\end{flushright}
Here we have studied the sensitivity of the order of polynomial approximation of the Airy's infinite series taking polynomial orders $r^3 , r^4 , r^6 , r^7 , r^9 , r^{10}$  (as polynomial orders $r^{(2+3l)}$ with l=0,1,2 etc are absent in the Airy's function series).
Further, it is found that the infinite limit of integration in calculating $\xi(y)$  and its derivatives makes the result divergent. We take some reasonable cut-off limit $r_0$ of its integration. This will not sacrifice the nature and value of Airy's function and its derivatives, because, Airy's function falls very sharply and almost dies out with increasing r-value beyond $r=5 $(Table 1 and Fig. 1). \\

\begin{figure}
    \centering
    \subfigure[  Airy's function with r]
    {
        \includegraphics[width=3.0in]{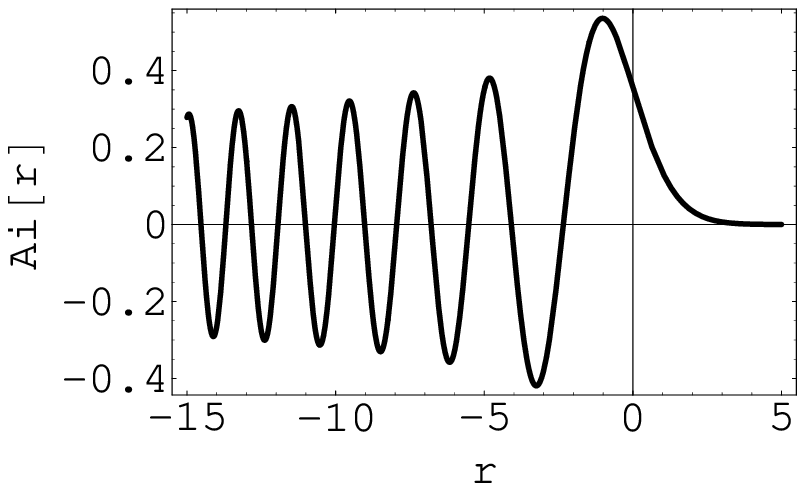}
        \label{fig:first_sub}
    }
        \subfigure[Airy's function with positive r]
    {
        \includegraphics[width=3.0in]{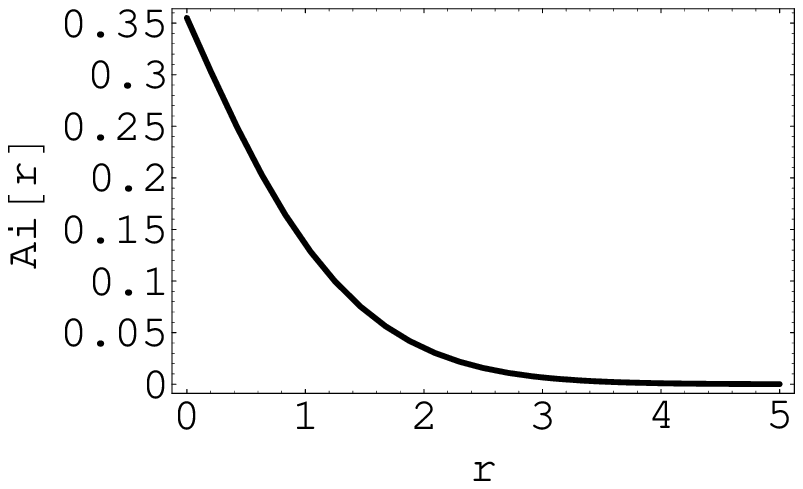}
        \label{fig:second_sub}
    }

    \caption{Variation of Airy's function with r}
    \label{fig:sample_subfigures}
\end{figure}

\begin{table}[ht]
\begin{center}
\caption{Values of Airy's function for some small positive r. }\label{cross}
\begin{tabular}{|l|l|l|l|}
  \hline
   r & Ai [r] & r & Ai [r] \\
  \hline \hline
  0.1 & 0.329 & 2.5 & 0.016 \\
  0.5 & 0.232 & 3.0 & 0.007 \\
  1.0 & 0.135 & 3.5 & 0.003 \\
  1.5 & 0.072 & 4.0 & 0.0009 \\
  2.0 & 0.035 & 4.5 & 0.0003 \\
  \hline
\end{tabular}
\end{center}
\end{table}

Also, the graph of normalization constants (N and $N^\prime$ )versus the cut-off to upper limit ( $r_0$ ) shows that N and $N^{\prime}$ values decrease with increase in $r_0$ ( Fig. 2 ).

\begin{figure}
    \centering
    \subfigure[  N vs $ r_0 $]
    {
        \includegraphics[width=3.0in]{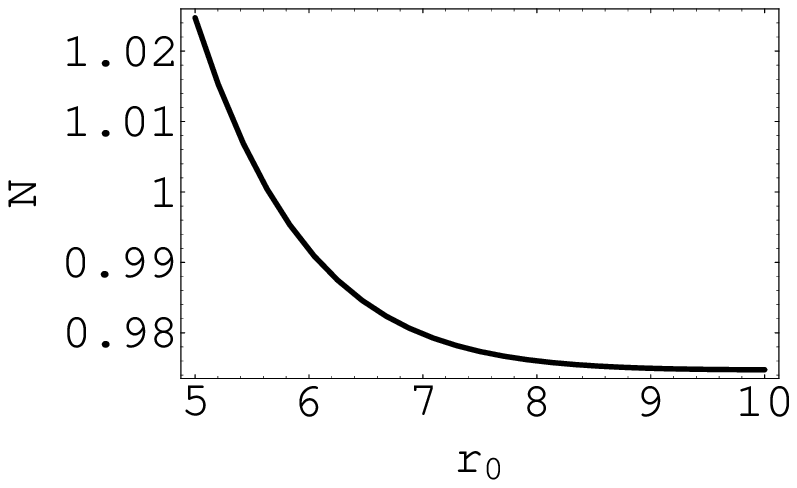}
        \label{fig:first_sub}
    }
        \subfigure[$ N^{\prime}$ vs $ r_0$]
    {
        \includegraphics[width=3.0in]{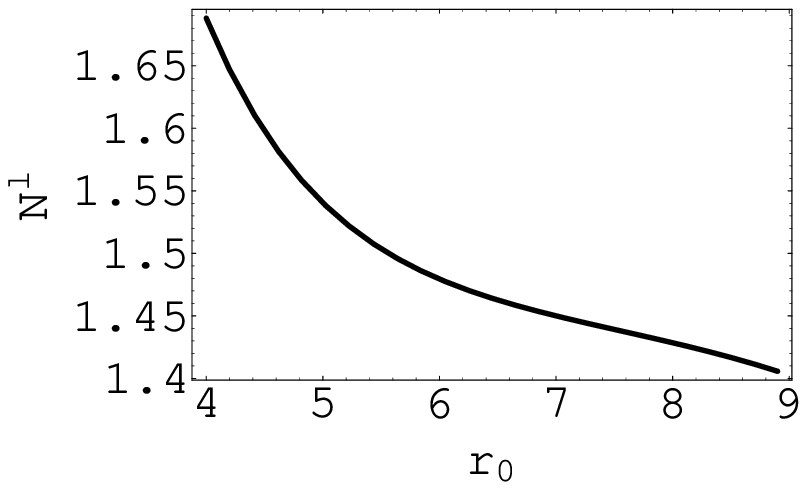}
        \label{fig:second_sub}
    }

    \caption{Variation of nornalisation constants with cut-off}
    \label{fig:sample_subfigures}
\end{figure}

Also, the graphs of $\rho^2$ vs $r_0$  and C vs $r_0$ ( Fig 3 ) confirm that beyond $r_0$ = 9, $\rho^2$ and C values rise steeply as compared to the result of Table 2. \\

\begin{figure}
    \centering
    \subfigure[$\rho^2 $ vs $r_0$]
    {
        \includegraphics[width=3.0in]{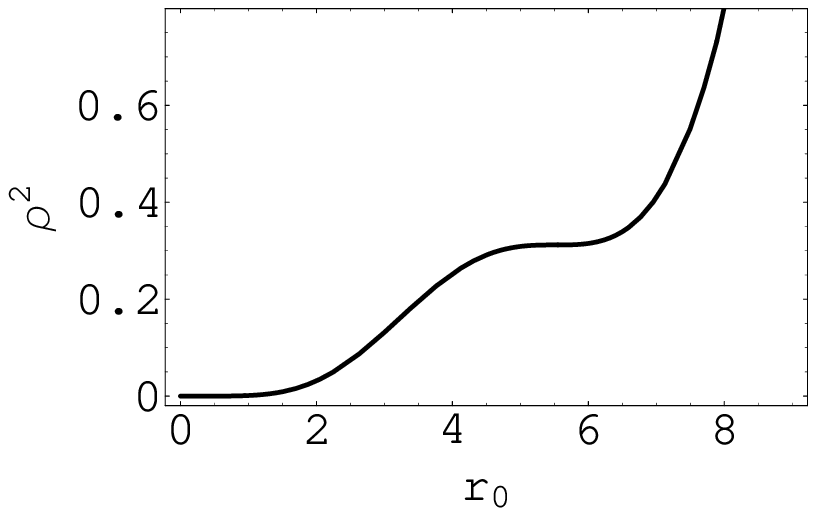}
        \label{fig:first_sub}
    }
        \subfigure[C vs $r_0$ ]
    {
        \includegraphics[width=3.0in]{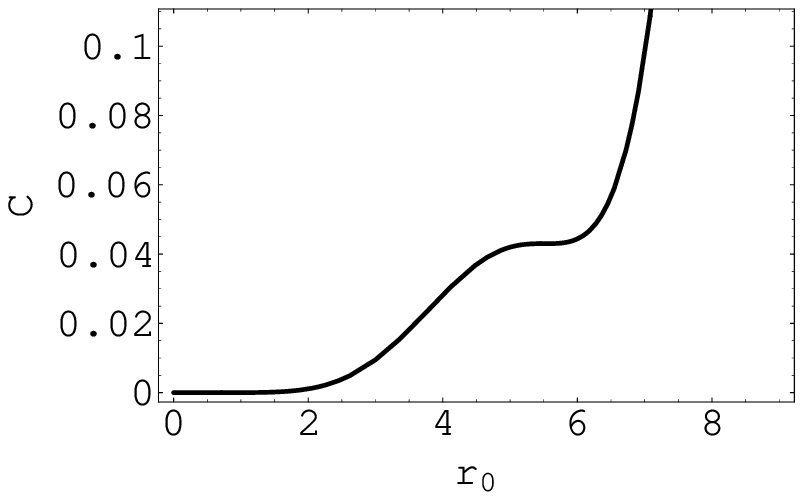}
        \label{fig:second_sub}
    }

\caption{Variation of slope and curvature with cut-off for Airy order $\sim r^{10}$ }
\label{fig:sample_subfigures}
\end{figure}

\begin{table}[ht]
\begin{center}
\caption{Results of slope and curvature of $\xi(y$) in different models and collaborations.}\label{cross}
\begin{tabular}{|c|r|r|}
  \hline
Model / collaboration &	Value of slope & Value of curvature \\
\hline \hline
 Ref [6] & 0.7936 & 0.0008 \\
 Le Youanc et al [24] & $\geq 0.75$ & $\geq 0.47$ \\
 Skryme Model [25] & 1.3 & 0.85 \\
 Neubert [26] & 0.82  0.09 & -- \\
 UK QCD Collab. [27]  & 0.83 & -- \\
 CLEO [28,29] & 1.67 & -- \\
 BELLE  [30] & 1.35 & -- \\
HFAG [31] & 1.17 $\pm 0.05$ & -- \\
Huang [32] & 1.35 $\pm 0.12 $ & -- \\
\hline
\end{tabular}
\end{center}
\end{table}

Upon this consideration, we have explored the $\xi(y)$  and its derivatives for different orders of polynomial approximation of Airy's function both for unperturbed wave function (Table 3) and total wave function (with relativistic effect) (Table 4), taking different cut-off values ranging from $r_0$ $=$ 5 to $r_0$ $=$ 9 , for D meson taking the input value $\alpha_s=0.22$ [22].

\begin{table}[ht]
\begin{center}
\caption{Result with unperturbed wave-function with $r_0$ = 5,7,9 $GeV^{-1}$.}\label{cross}
\begin{tabular}{|c|ccc|ccc|ccc|}

  \hline
  O(r)in &  &$r_0=5 $&  &  &$r_0=7 $&  &  &$r_0=9$ &  \\

   $A_i$ & N & $\rho^2$ & C & N & $\rho^2$ & C & N & $\rho^2$ & C \\
  \hline \hline
  $r^3$ & 0.9307 & 0.5053 & 0.0872 & 0.8864 & 0.7595 & 0.2213 & 0.8656 & 0.9857 & 0.4581 \\
  $r^4$ & 1.0658 & 0.6712 & 0.1123 & 1.0121 & 0.8625 & 0.2195	 & 1.0118	 & 0.8658	 & 0.2229 \\
  $r^6$ & 1.0374 & 0.6414 & 0.1066 & 0.9865 & 0.8343 & 0.2161	 & 0.9835	 & 0.8634	 & 0.2468 \\
  $r^7$ & 1.0201 & 0.6221 & 0.1031 & 0.9720 & 0.8078 & 0.2071	 & 0.9711	 & 0.8160	 & 0.2153 \\
  $r^9$ & 1.0235 & 0.6258 & 0.1037 & 0.9749 & 0.8128 & 0.2086	& 0.9737	 & 0.8236	 & 0.2196 \\
  $r^{10}$ &1.0248 & 0.6274 & 0.1040 & 0.9760 & 0.8144 & 0.2090	 & 0.9750	 & 0.8242	&0.2189 \\
  \hline

\end{tabular}
\end{center}
\end{table}

\begin{table}[ht]
\begin{center}
\caption{Result with total wave-function ( with relativistic effect) taking $r_0$ = 5,7,9 $GeV^{-1}$.}\label{centre}
\begin{tabular}{|c|ccc|ccc|ccc|}

  \hline
  O(r)in &  &$r_0=5 $&  &  &$r_0=7 $&  &  &$r_0=9$ &  \\

   $A_i$ & N & $\rho^2$ & C & N & $\rho^2$ & C & N & $\rho^2$ & C \\
  \hline \hline
  $r^3$ &1.5927&	0.5149	&0.0661&	1.5658 &	0.6125&	0.1269&	1.4985&	0.9681	&0.4365 \\
  $r^4$ & 1.8653&	0.6300	&0.0869	&1.8104	&0.7942	&0.1932	&1.4375	&2.4989&	1.8879 \\
  $r^6$ &1.8703&	0.6432	&0.0878	&1.8162	&0.8040	&0.1921	&1.5311	&2.0925&	1.4440 \\
  $r^7$ &1.8471&	0.6313	&0.0858	&1.7946	&0.7899	&0.1884	&1.4769	&2.2604	&1.6355 \\
  $r^9$ &1.8518&	0.6340	&0.0862	&1.7989	&0.7931	&0.1892	&1.4871	&2.2294	&1.5990 \\
  $r^{10}$ &1.8531&	0.6348	&0.0863	&1.8000	&0.7940	&0.1895	&1.4847	&2.2462	&1.6169 \\
  \hline

\end{tabular}
\end{center}
\end{table}

\begin{figure}
    \centering
    \subfigure[  IWF vs y for diff. cut-off(with Airy order $r^3$)]
    {
        \includegraphics[width=3.0in]{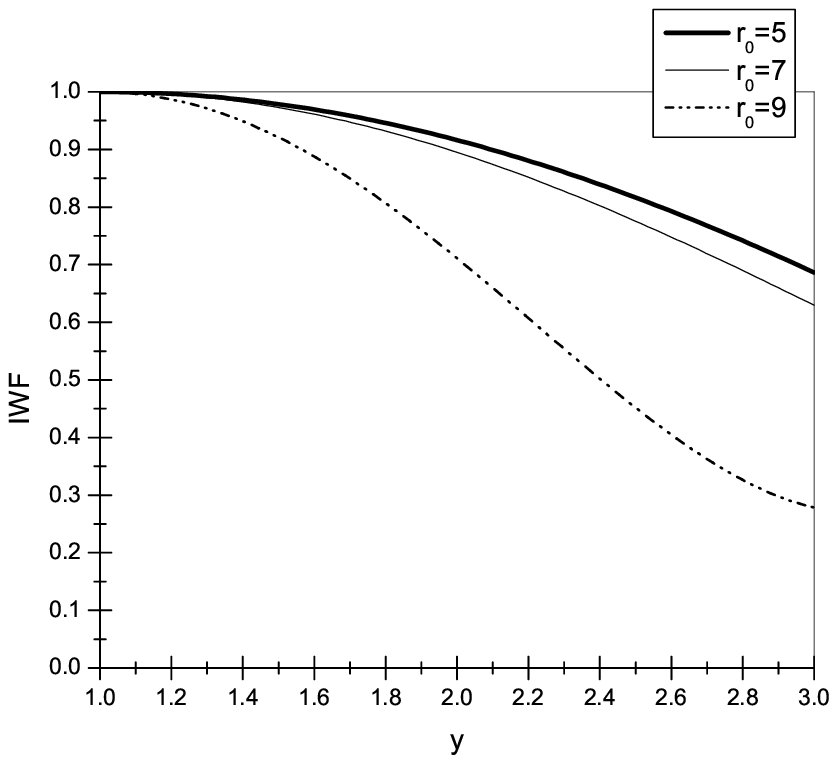}
        \label{fig:first_sub}
    }
        \subfigure[IWF vs y for diff. Airy-order ( with $r_0 = 5$) ]
    {
        \includegraphics[width=3.0in]{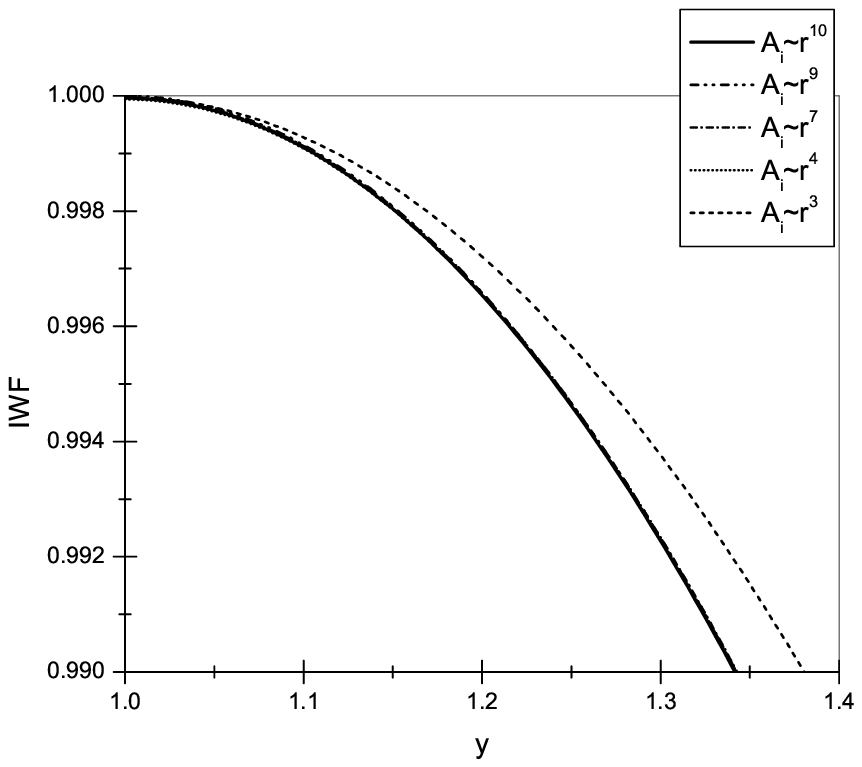}
        \label{fig:second_sub}
    }
\caption{Variation of IWF with y}
\label{fig:sample_subfigures}
\end{figure}

Regarding sensitivity of the order of polynomial in infinite Airy's function, the result for $\rho^2$ and C do not differ much upon variation of order of polynomial in Airy's function. For a given Airy order, with increase in cut-off value, $\rho^2$ and C values increases steadily, whereas for a given cut-off value, $\rho^2$ and C values do not differ much upon variation of order of polynomial in Airy's function from $r^4$ to $r^{10}$. However, the results show closer resemblance to recent result of $\rho^2$ = 1.17 [31] for our Airy-order $r^3$ up to cut-off value $r_0$ = 10. For such specific range and order, our result shows improvement upon the result of ref [6]. At cut-off value higher than $r_0$ = 9, the results jumps to higher values than our expectations.\\
 The variation of $\xi(y)$ with y for different cut-off values $r_0$ ,with Airy order $r^{10}$ , is shown in figure 4(a) and the variation of $\xi(y)$ with y for different Airy order at cut-off value $r_0$ = 5 is shown in Fig. 4(b). In 4(b), graphs of $\xi(y)$ vs  y overlaps for Airy orders $r^4$ to $r^{10}$ , whereas the graph for Airy order $r^3$ shows a slight deviation.
Thus, these graphs are in agreement with our expectations [8]. The graphs of $\xi(y)$  with y for different Airy's function order invariably start at (1,1) and almost follow the same pattern and show very small deviation with change in Airy order. It confirms the fact that boundary condition for zero recoil $(\xi(1)=1)$ is maintained all through, with different polynomial orders of Airy's function and for different cut-off values. \\

\section{Conclusion and remarks:}
We have found that cutting off the upper limit of integrations in $\xi(y)$ and its derivatives to some reasonable point does not upset the result, rather it almost conforms to the experimental expectations.
Also, for each value of cut-off $r_0$ , we have considered the asymptotic form of the Airy's function taking limits of integration from $r_0$  to $\infty$ .
\begin{equation}
A_i[\varrho]_{asympt} \sim \frac{ \exp{(-\frac{2}{3}\varrho^{3/2}})}{2\sqrt{\pi}\varrho^{1/4}}
\end{equation}
With this asymptotic form we have also calculated the derivatives of $\xi(y)$. Such analysis shows that very small values of  $\rho^2$ and C result [Table 5], taking this asymptotic form of Airy's function. Thus, the margin of error in the results of $\rho^{2}$ and $C$ due to cutting off the upper limit of integration to some reasonable value is negligible , as is evident from these very small asymptotic values.\\
\begin{table}[ht]
\begin{center}
\caption{Values of  $\rho^2$ and C with asymptotic form of Airy's function.}\label{centre}
\begin{tabular}{|c|c|c|}
  \hline
  $r_0$ value & $\rho^2$ (asymptotic ) & C(asymptotic)\\
  \hline \hline
   5 & $4.6 \times  10^{-9}$ & $1.6 \times 10^{-9}$ \\
   \hline
   6 & $5.027 \times  10^{-11}$ &	$2.464 \times 10^{-11}$ \\
   \hline
   7& $3.56 \times  10^{-13}$ &	$2.345 \times 10^{-13}$ \\
   \hline
   8&  $1.695 \times  10^{-15}$ 	& $ 7.028 \times 10^{-15}$ \\
   \hline
   9&  $5.248 \times  10^{-18}$	& $6.597 \times 10^{-15}$ \\
   \hline
   10& $2.92 \times  10^{-19}$	& $2.78 \times  10^{-19}$ \\
    \hline
\end{tabular}
\end{center}
\end{table}
\begin{table}[h]
\begin{center}
\caption{Value of cut-off $r_0$ for different Airy order matching expectation of  $\rho^2$.}\label{centre}
\begin{tabular}{|cc|}
  \hline
  $\rho^2=1.17 GeV^{-1}(ref [31])$ & \\
  \hline \hline
  Airy Order &	$r_0$ value \\
  \hline \hline
  $r^3$ &  8.896 \\
  $r^4$ & 7.915 \\
  $r^6$ & 7.975 \\
  $r^7$ & 7.942 \\
  $r^9$ & 7.939 \\
  $r^{10} $ &7.932 \\
    \hline
\end{tabular}
\end{center}
\end{table}

Let us also comment on the result of Ref [6]. The result of ref [6], which is for Airy order $r^3$, matches with our cut-off value $r_0 =7.95 \; GeV^{-1}$ for the same Airy order in our calculation. However, the wave function in ref [6] is not found to satisfy the zero recoil condition of   IWF. \\
To conclude, we also study the compatibility our potential model with the recent results of Heavy Flavour Averaging Group [31]. Taking the result of $\rho^2$, we fix the value of cut-off for different orders of polynomial in Airy function (Table 6). It indicates that, the range of cut-off  value $r_0 = 7 \; GeV^{-1}$ to $r_0 = 9 \;  GeV^{-1}$  matches the expectations of ref [31]. \\
Further, regarding uncertainty of our results, we would like to mention here that the value of confinement parameter $b$ has been taken as 0.183 $GeV^{2} $ from charmanium spectroscopy [9]. However, for B -sector mesons,the value of b might be different. Also, there is no standard value of the constant c in QCD potential. In our calculation, we have taken c to be 1 GeV [5] to make it compatible with the masses of the mesons. Here may lie some margin of uncertainty in our result. \\
Lastly, we would also like to comment on the limitations of the present approach.\\
(i)	The present approach falls short of theoretically more sophisticated approach of lattice QCD, as far as numerical accuracy is concerned. \\
(ii)	Similarly, numerical solution of Schrodinger equation with the specific potential also gives more accurate result than the present one. However, this  approach appears to lack physical insight into the problem unlike the relatively crude potential model approach pursued here. \\
(iii)	We would also like to comment on the limitation of Airy's function as a realistic meson wave function and its possible ways-out. The Airy's infinite series, by definition, cannot be normalised. We have therefore introduced reasonable cut-off value so as to conform it to the experimentally measurable quantities like slope of Isgur-Wise function for heavy-light meson up to a given order of polynomial approximation  of the Airy's function. This model is therefore an improvement of the earlier work in the subject ( ref.[6]).\\
(iv)	Another limitation of the present formalism is that the perturbed wave function $\Psi^{\prime}(r)$   is up to Airy order $r^3$, as in ref [6]. Although, the total wave function  contains infinite Airy series in terms of unperturbed wave function, this above mentioned limitation may have some effect on the result. Improvement of the formalism considering higher polynomial orders of Airy's function in $\Psi^{\prime}(r)$  is under consideration.

\paragraph{Acknowledgement :\\ }
\begin{flushleft}
\emph{One of the authors ( SR ) acknowledges the support of University Grants Commission in terms of fellowship under FDP scheme to pursue research work at Gauhati University. Both the authors extend heartiest thanks to the authority of Pandu College, Guwahati for providing necessary facilities.}
\end{flushleft}

\end{document}